%% file: evalproj.tex
\documentclass{lib/egpubl}
\EGyear{~}   
\ConfYear{~} 
\ConfEditors{~} 
\ConfEditorStrg{~} 
\ConfName{~} 
\volume{34}     
\issue{3}       
\pdfSubject{}

\makeatletter
\def\NoIssuePaper{%
\EGpagenumber%
\copyrightTextTitPag{}
\def\@oddfoot{}
\def\@evenfoot{}
\def\ps@titlepage{\let\@mkboth\@gobbletwo
 \def\@oddhead{}%
}
 \def\@oddfoot{}
 \let\@evenhead=\@oddhead
 \let\sectionmark=\EmptySectionmark
 \let\subsectionmark=\EmptySubsectionmark
}
\makeatother
\NoIssuePaper
\electronicVersion 

\ifpdf \usepackage[pdftex]{graphicx} \pdfcompresslevel=9
\else \usepackage[dvips]{graphicx} \fi

\usepackage{xspace}
\usepackage{amsfonts}
\usepackage{amsmath}
\usepackage[hang,raggedright]{subfigure}
\usepackage{enumitem}
\usepackage{hyperref}
\usepackage{color}
\usepackage{mathptmx}
\usepackage{comment}

\graphicspath{{img/}}

\input{tex/figures}

\usepackage{t1enc, lib/dfadobe}

\usepackage{lib/egweblnk}
\usepackage{cite}

\title[Visualizing Dimensionality Reduction Artifacts: An Evaluation]
      {Visualizing Dimensionality Reduction Artifacts: \\ An Evaluation}

\author[N. Heulot, J.-D. Fekete \& M. Aupetit]{
  Nicolas Heulot$^{1,2}$, Jean-Daniel Fekete$^1$, Michael Aupetit$^2$\\
  $^1$Inria, France, $^2$CEA, France\\
  May 2015}

\teaser{\centering
	\includegraphics[width=8cm, height=5.0cm]{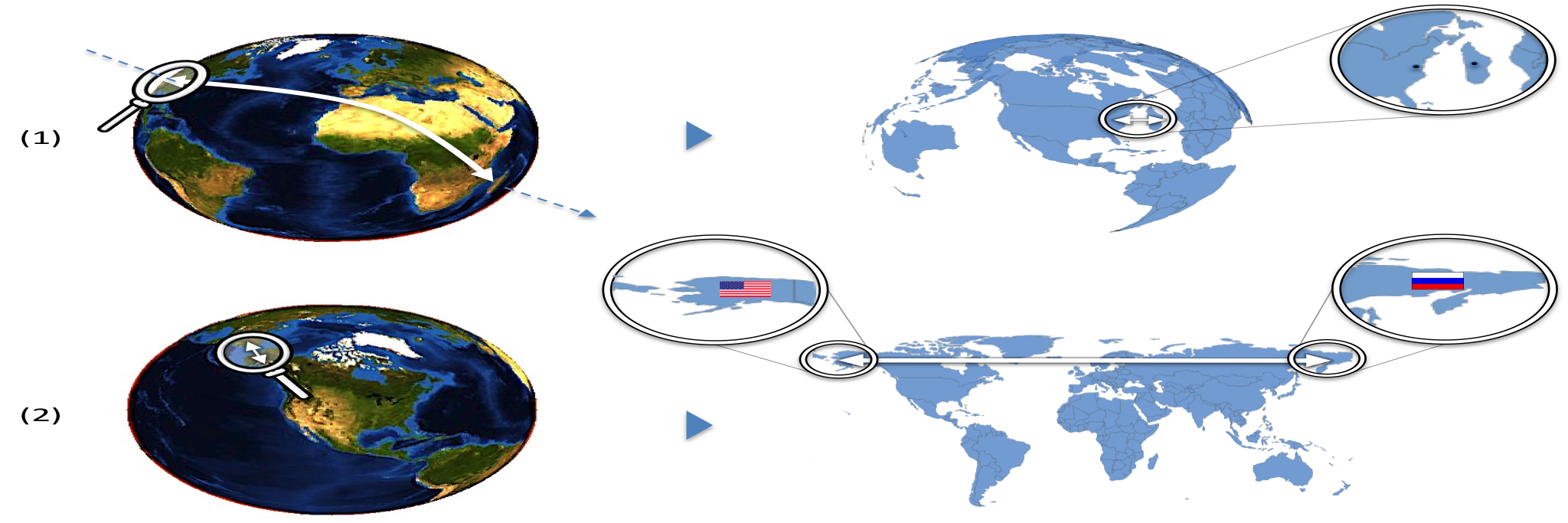} 
	\caption{Multidimensional projection artifacts are identical
          to geographic projections artifacts when we project the Earth on a 2D plane:
		(1) The orthogonal projection exhibits false neighborhoods: here, Washington appears right next to Antananarivo.
		(2) The Mercator projection exhibits tears: the Earth has been unfolded along the equator, leading to display Alaska far away from Russia.
		These artifacts always exist when projecting high-dimensional data with any dimensionality reduction technique.
	}
	\label{fig:teaser}
}

\begin{document}

\maketitle

\begin{abstract}
\input{tex/abstract}

\end{abstract}

\input{tex/introduction}
\input{tex/background}

\input{tex/evaluation}

\input{tex/results}

\input{tex/discussion}

\input{tex/conclusion}


\bibliographystyle{lib/eg-alpha-doi}

\bibliography{evalproj}


\end{document}

%% file: tex/figures.tex
\newcommand{\eg}{\textit{e.\,g.}\xspace}
\newcommand{\ie}{\textit{i.\,e.}\xspace}

\def\displayFigures{0}

\newcommand{\figureArtifacts}{
	\begin{figure}
	\begin{center}
	\includegraphics[width=6.5cm, height=3.5cm]{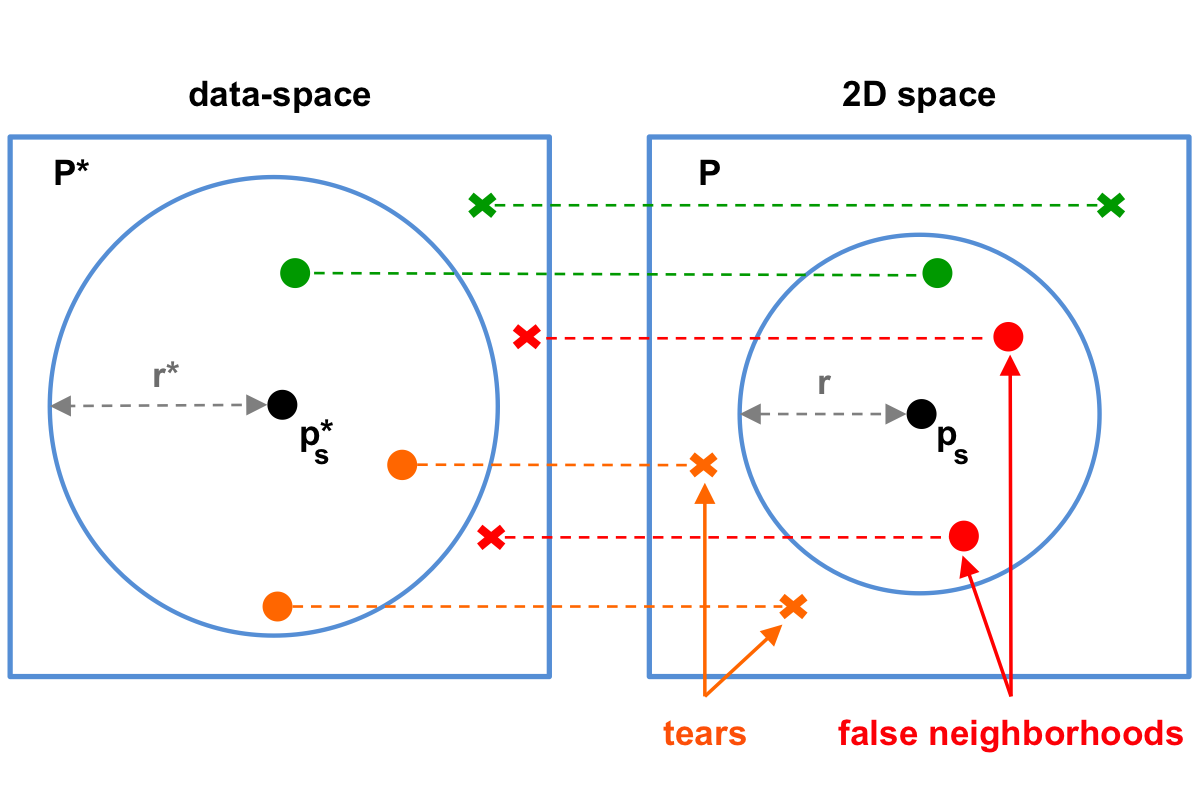}
	\caption{Model characterizing the different types of points relative to the reference point (in black). 
		The neighborhood of the reference in the data-space is represented by a sphere of radius $r^*$ centered in the reference $p_s^*$.
		And in the same way, its neighborhood in the 2D space is represented by a disc of radius $r$ centered in $p_s$.
		Points in the neighborhood of the reference are represented by a circle and the other by a cross.
		The points are either topological artifacts, \ie false neighborhoods (in red) and tears (in orange), or well placed points (in green).
	}
	\label{fig:distortions}
	\end{center}
	\end{figure}
}

\newcommand{\figureTasks}{
	\begin{figure*}
	\begin{center}
	\includegraphics[width=16cm, height=5.0cm]{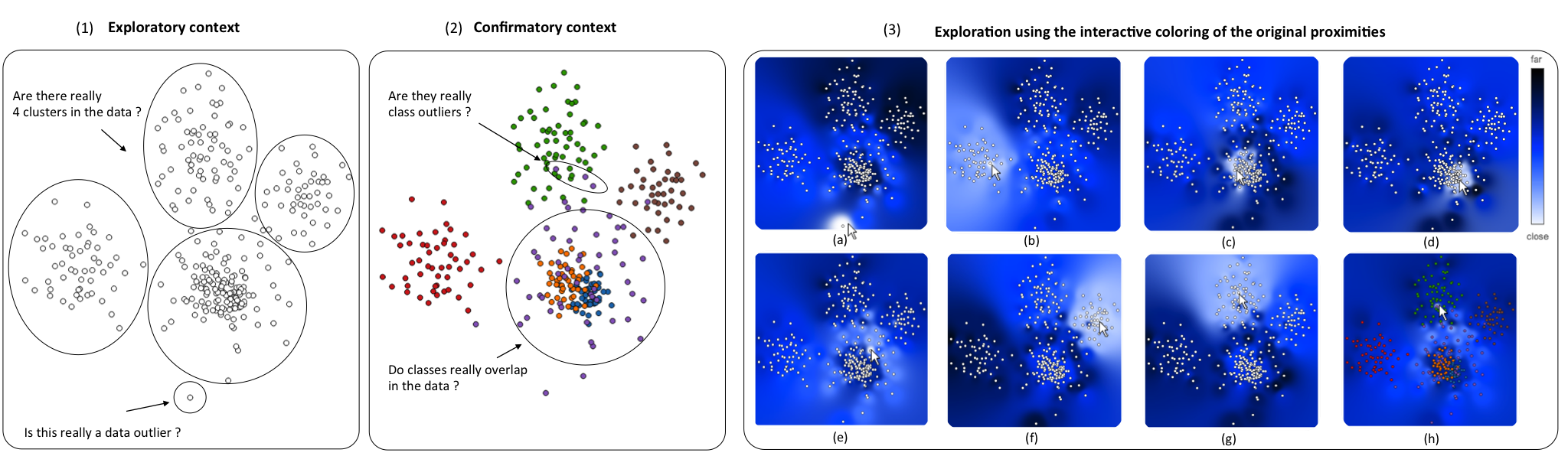} 
	\caption{Main questions involved by the visual analysis of projections in exploratory and confirmatory contexts (1,2).
	Exploring the projection with the interactive coloring of the original dissimilarities gives an answer to each question (3).
	Indeed, the gaps in the projection colorings (from white to blue) show that there are six underlying clusters in the data (b,c,e,f,g,h).
	We also observe that the point at the bottom of the projection (1) seems to be actually a data outlier (a) and that the three points at the top of the projection (2) are not actually class outliers (h).
	The projection was obtained from the PCA of a synthetic dataset composed of 6 clusters generated from Gaussian distributions randomly centered in a 10-dimensional space.
	The reference point of the interactive coloring is the point with the whitest color, and is also the closest point from the mouse cursor. The whiter the color, the lower the dissimilarity to the reference in the data-space.
	}
	\label{fig:tasks}
	\end{center}
	\end{figure*}
}

\newcommand{\figureDatasetsDetails}{
	\begin{figure}
	\begin{center}
	 \scriptsize
	\begin{tabular}{|l|c|c|c|c|c|}
	  \hline
	  \textbf{Name} & \textbf{Dimensions}  & \textbf{Instances} & \textbf{Clusters} & \textbf{MDS} \\
	  \hline
	  gaussians & 100 & 300 & 6 & NerV \\
	  \hline
	  noisy gaussians & 100 & 240 & 6 & NerV \\
	  \hline
	  pen-digits & 16 & 250 & 5 & classical \\
	  \hline
	  opt-digits & 64 & 300 & 6 & NerV \\
	  \hline
	  isolet & 617 & 200 & 4 & classical \\
	  \hline
	  letters & 16 & 180 & 6 & classical \\
	  \hline
	  faces & 960 & 128 & 4 & NerV \\
	  \hline
	  teapot & 192 & 365 & 2 & NerV \\
	  \hline 
	\end{tabular}
	\caption{Summary of the characteristics of each dataset used in the experiment.}
	\label{fig:dataset_summary}
	\end{center}
	\end{figure}
}

\newcommand{\figureProxiViz}{
	\begin{figure}
	\begin{center}
	\begin{tabular}{ c c }
	  \subfigure[Bare projection]{ \colorbox{white}{\includegraphics[width=3.0cm, height=3.0cm]{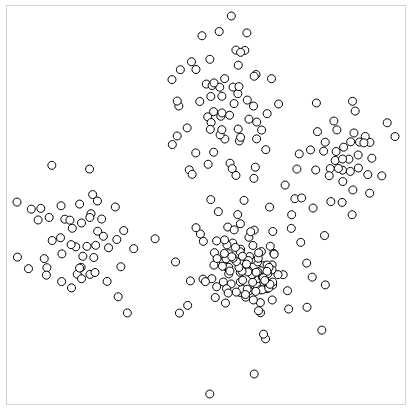}}\label{fig:bare_projection} }
	  & \subfigure[ProxiViz]{ \colorbox{white}{\includegraphics[width=3.5cm, height=3.0cm]{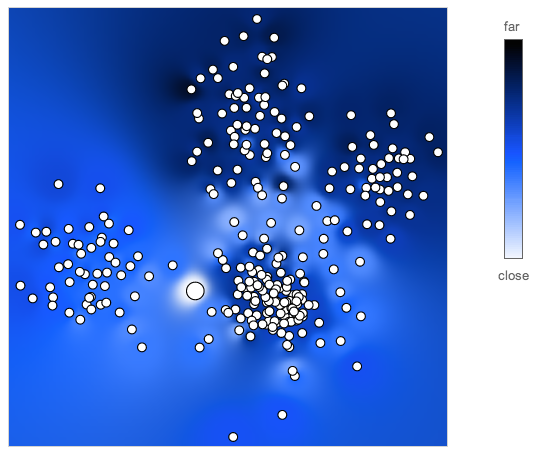}}\label{fig:proxiViz2} }
	\end{tabular}
	\caption{Projection and ProxiViz as used in the experiment.
	The colorscale was shown to help analyzing the coloring. 
	The gaps between colors intensity on the projection indicate gaps of dissimilarities in the data-space. 
	}
	\label{fig:proxiviz}
	\end{center}
	\end{figure}
}

\newcommand{\figureDatasets}{
	\newlength\imgWidthDatasets
	\newlength\imgHeightDatasets
	\imgWidthDatasets=3.5cm
	\imgHeightDatasets=3.0cm
	\begin{figure*}
	\begin{center}
	\begin{tabular}{ c c c c }
	  \subfigure[faces]{ \colorbox{white}{\includegraphics[width=\imgWidthDatasets, height=\imgHeightDatasets]{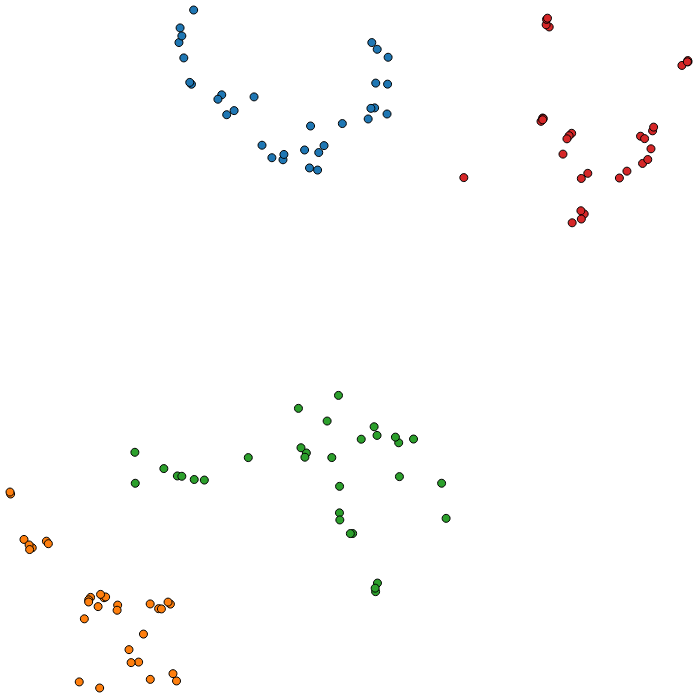}}\label{fig:dataset1} }
	  & \subfigure[gaussians]{ \colorbox{white}{\includegraphics[width=\imgWidthDatasets, height=\imgHeightDatasets]{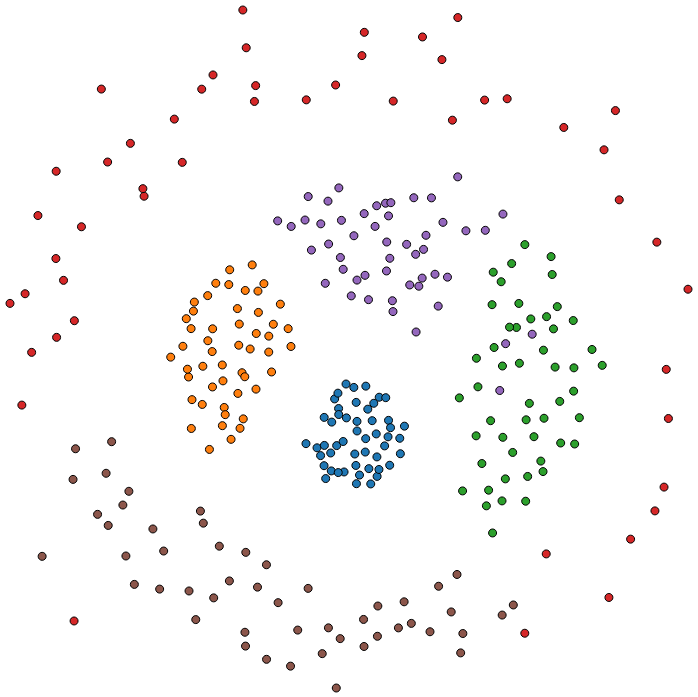}}\label{fig:dataset2} }
	  & \subfigure[pen digits]{ \colorbox{white}{\includegraphics[width=\imgWidthDatasets, height=\imgHeightDatasets]{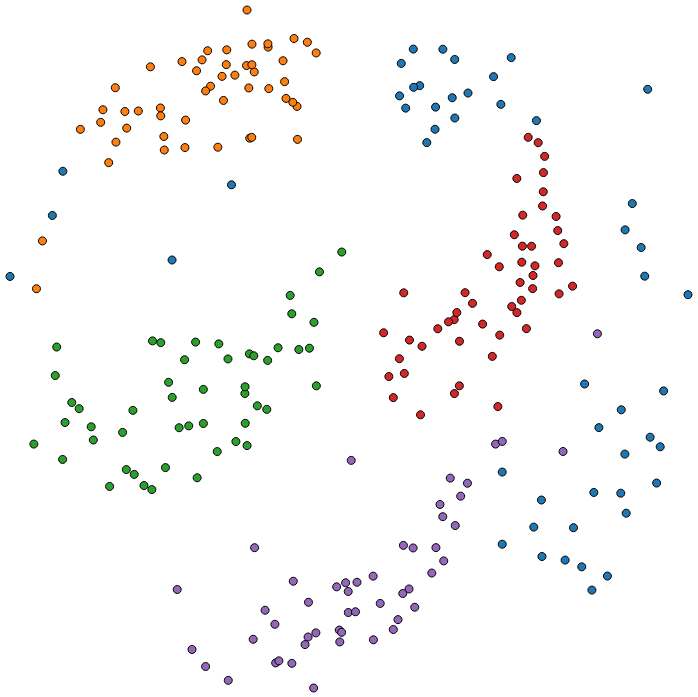}}\label{fig:dataset3} }
	  & \subfigure[opt digits]{ \colorbox{white}{\includegraphics[width=\imgWidthDatasets, height=\imgHeightDatasets]{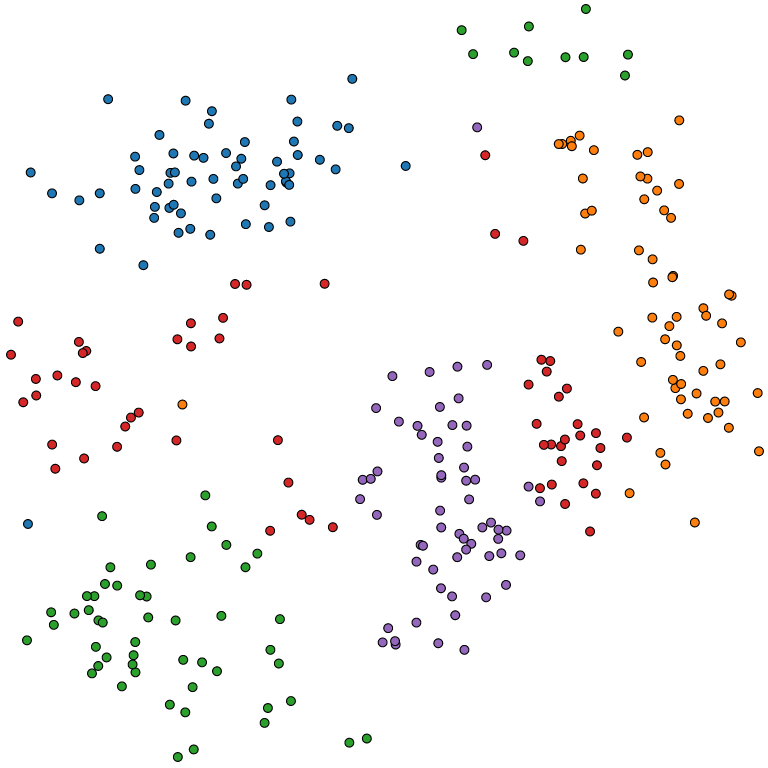}}\label{fig:dataset4} } \\
	  \subfigure[noisy gaussians]{ \colorbox{white}{\includegraphics[width=\imgWidthDatasets, height=\imgHeightDatasets]{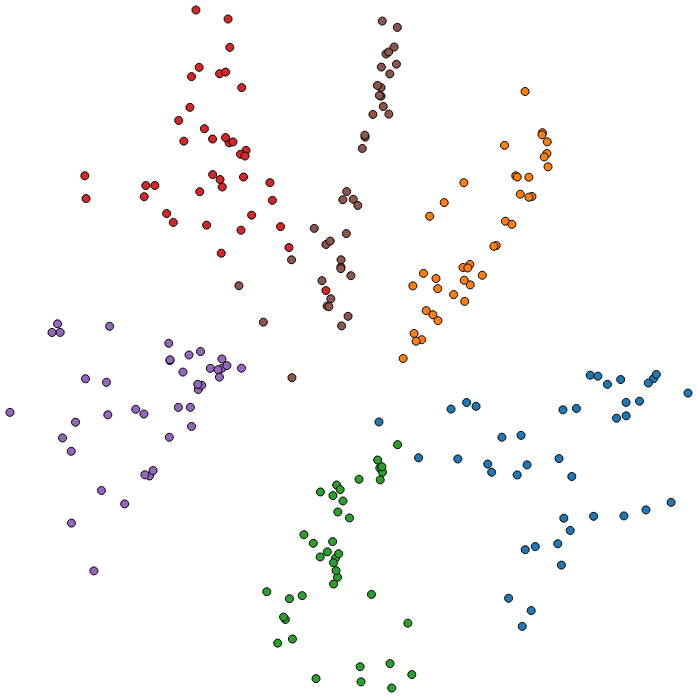}}\label{fig:dataset5} }
	  & \subfigure[isolet]{ \colorbox{white}{\includegraphics[width=\imgWidthDatasets, height=\imgHeightDatasets]{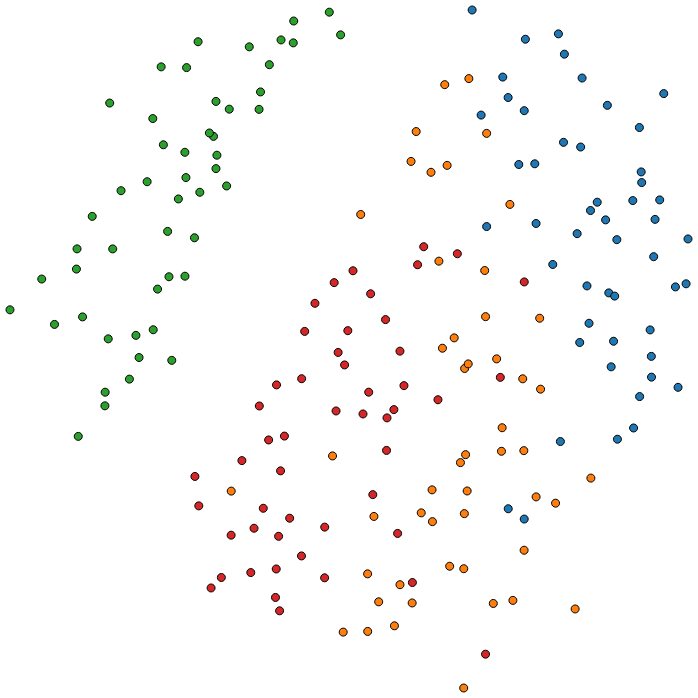}}\label{fig:dataset6} }
	  & \subfigure[teapot]{ \colorbox{white}{\includegraphics[width=\imgWidthDatasets, height=\imgHeightDatasets]{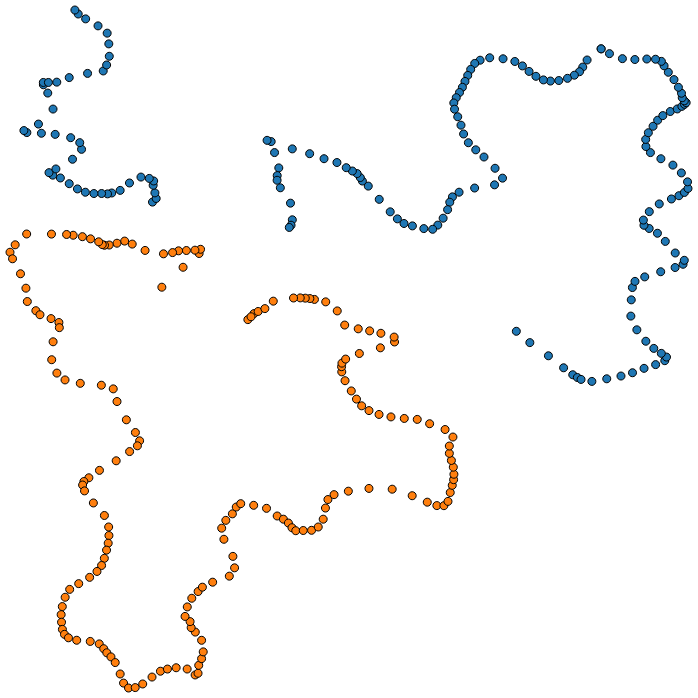}}\label{fig:dataset7} }
	  & \subfigure[letter recognition]{ \colorbox{white}{\includegraphics[width=\imgWidthDatasets, height=\imgHeightDatasets]{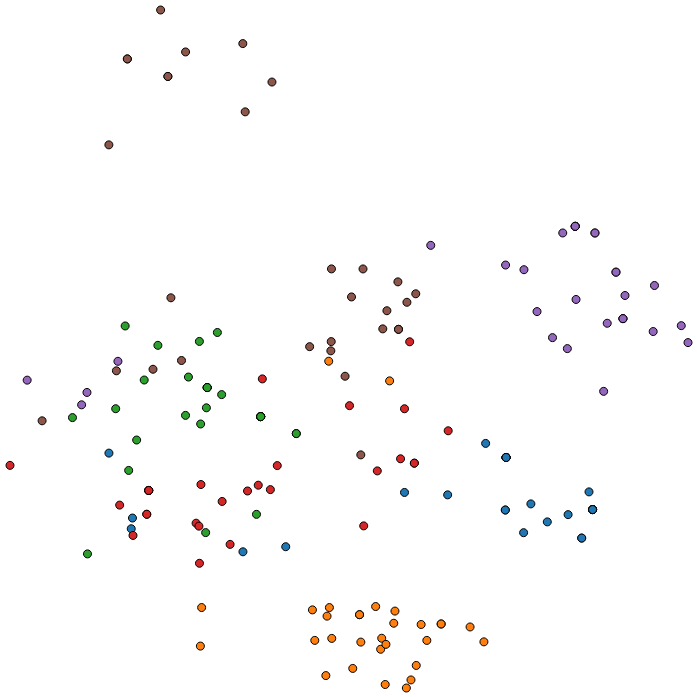}}\label{fig:dataset8} }
	\end{tabular}
	\caption{Projections of the different datasets used in the experiment. The colors represent class labels. Datasets 1-4 were considered as easy for a visual clustering task and datasets 5-8 were considered more difficult to analyze as the 2D clusters are less separated. Labels were indicated only for the class outliers validation task.}
	\label{fig:evalproj2}
	\end{center}
	\end{figure*}
}

\newcommand{\figureResultsCorrectness}{
	\begin{figure*}
	\begin{center}
	\includegraphics[width=16.5cm, height=11.0cm]{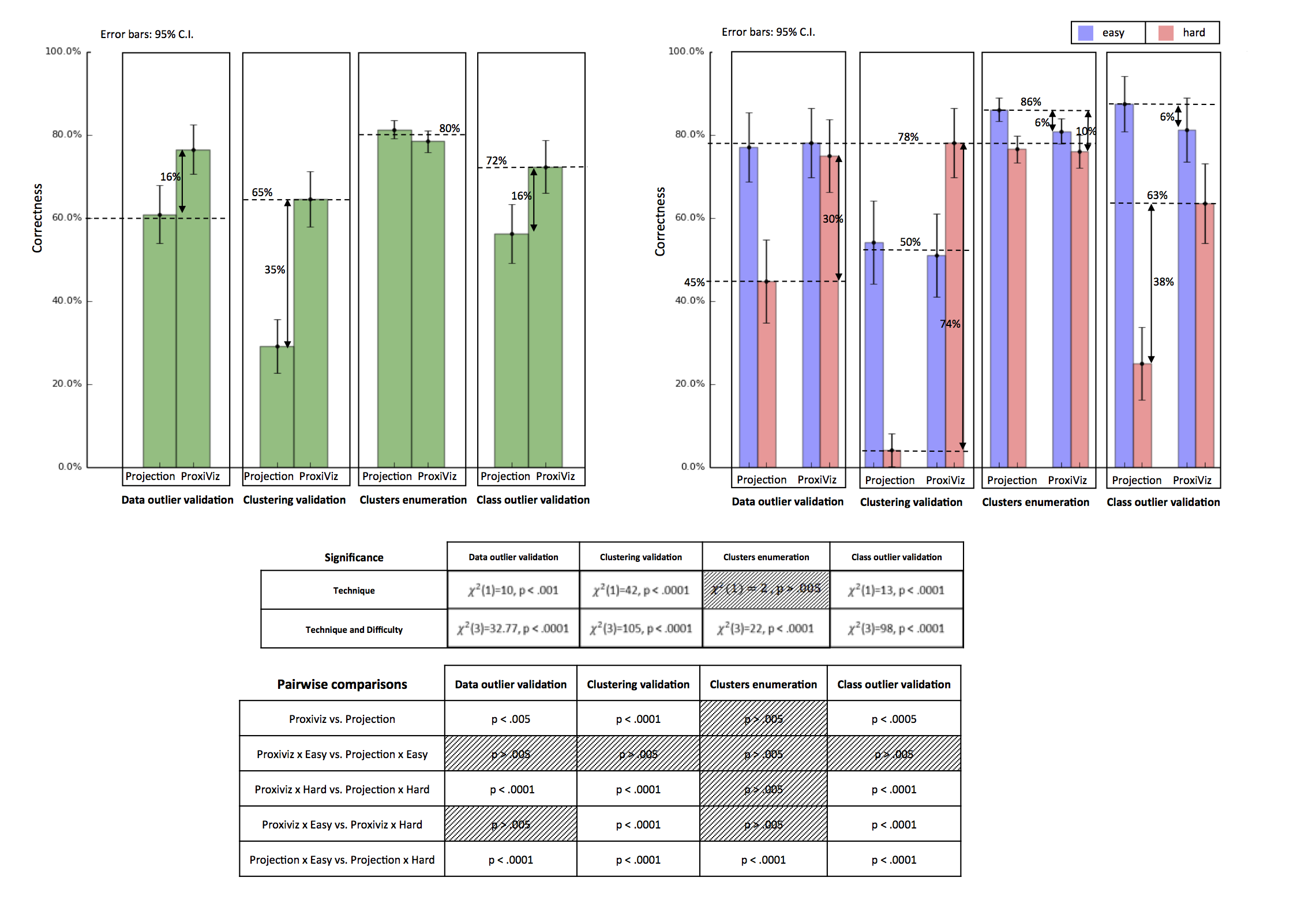} 
	\caption{Results of average correctness (in \%), significance between factors given by Friedman's test, and pairwise Wilcoxon signed rank test results.}
	\label{fig:results1}
	\end{center}
	\end{figure*}
}

\newcommand{\figureResultsTime}{
	\begin{figure*}
	\begin{center}
	\includegraphics[width=16.5cm, height=7.5cm]{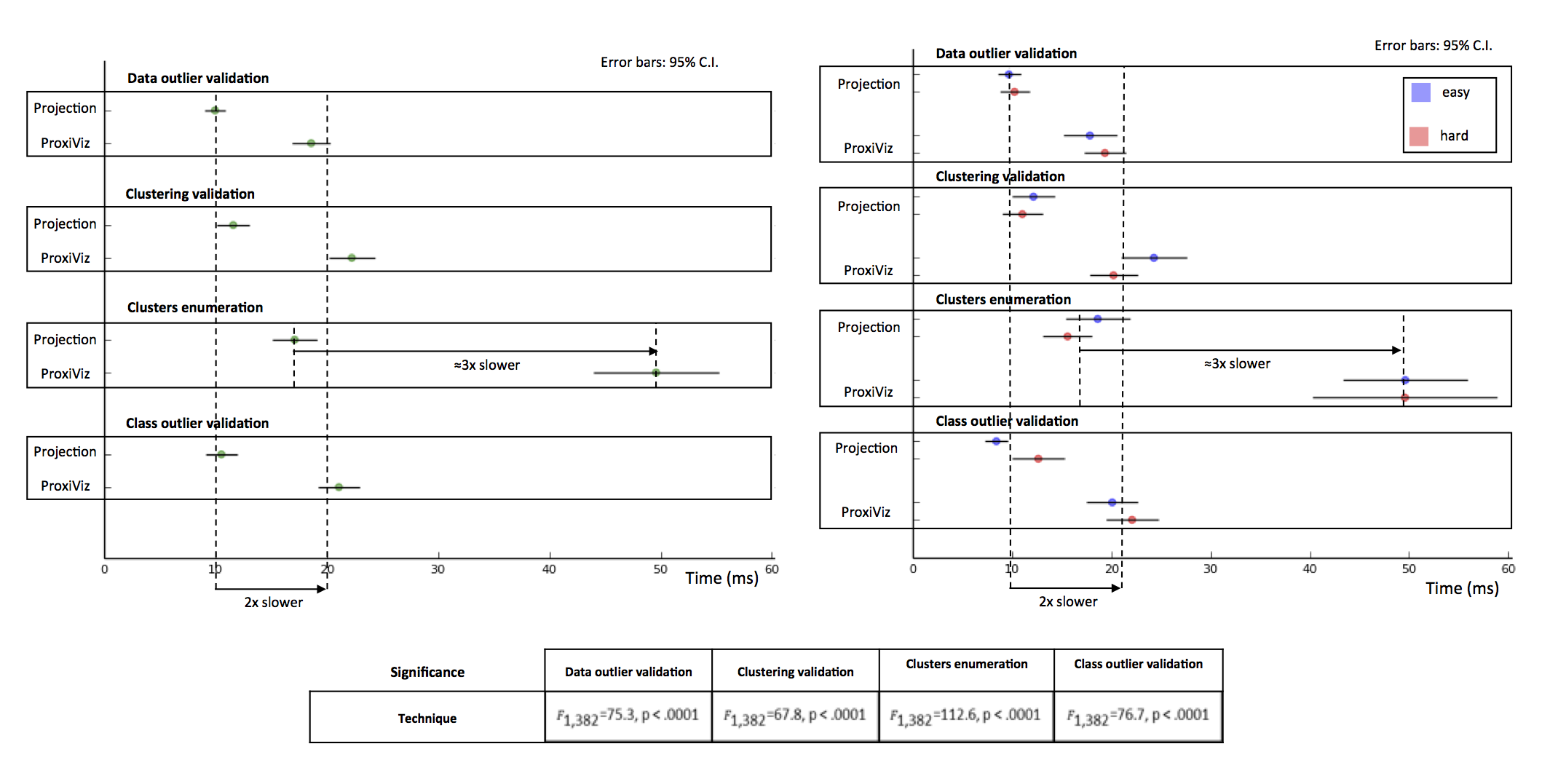} 
	\caption{Results of average response time (in seconds) and significance between factors given by ANOVA.}
	\label{fig:results2}
	\end{center}
	\end{figure*}
}

%% file: tex/abstract.tex
Multidimensional scaling allows visualizing high-dimensional data as
2D maps with the premise that insights in 2D reveal valid information
in high-dimensions, but the resulting projections always suffer from
artifacts such as false neighborhoods and tears.  These artifacts can
be revealed by interactively coloring the projection according to the
original dissimilarities relative to a reference item. However, it is not
clear if conveying these dissimilarities using color and displaying only
local information really helps to overcome the projections artifacts.
We conducted a controlled experiment to investigate the relevance of
this interactive technique using several datasets.  We
compared the bare projection with the interactive coloring of the
original dissimilarities on different visual analysis tasks involving
outliers and clusters.  Results indicate that the interactive coloring
is effective for local tasks as it is robust to projection artifacts
whereas using the bare projection alone is error prone.

%% file: tex/introduction.tex
\section{Introduction}

Multidimensional data is ubiquitous: all domains including sciences
and engineering need to analyze and make sense of it.  Several
visualization techniques are known to be effective for data up to
10-15 dimensions, such as scatterplot
matrices~\cite{cleveland1988dynamic} and parallel
coordinates~\cite{inselberg1985parallel}, but they become unusable
when the number of dimensions grows. Data then becomes
high-dimensional and fewer techniques are effective.  Multidimensional
scaling is one of them: it tackles the visualization problem by
summarizing many dimensions into a similarity matrix that is
visualized as a 2D scatterplot called a \emph{projection}: 2D distances
between points in the projection are meant to preserve the original
dissimilarities between data items in high-dimensions.

Projections have been mainly applied in domains where users are data
scientists who want to get insights about their high-dimensional data
\cite{brehmer2014visualizing}.  However, the most promising
applications might be in domains where users are not professional
analysts but need to make decisions about high-dimensional data.  The
assumption being that if visual patterns visible in 2D reveal patterns
in high-dimensions, then with only a short training, anyone should be
able to detect these patterns.

A large number of projection algorithms
\cite{maaten2008dimensionality} have been designed to effectively and
efficiently map a set of data items from a high-dimensional space
(\emph{data-space}) to a lower dimensional space (\emph{2D-space}),
while preserving as much information as possible.  However, as good as
these algorithms can be, the dimensionality reduction process
necessarily implies a loss of information that is materialized by
distortions such as topological artifacts 
(\autoref{fig:teaser}).  These artifacts interfere with the visual
analysis process and challenge the \emph{interpretation} and
\emph{trust} \cite{chuang2012designing} of projections.

Some existing measures quantify and can reveal artifacts, \eg through
coloring problematic points \cite{martins2014visual}; they can also
help choosing relevant projections through visual quality criteria
\cite{bertini2011quality}. However, these existing approaches provide
hints on possible artifacts but do not actually help overcome them for
analyzing data clustering.

Aupetit introduced a technique in \cite{aupetit2007visualizing} that
colors projections according to the original dissimilarities relative to a
reference item interactively selected by a pointer.  This technique
not only reveals artifacts but additionally shows information about
the data-space through the color channel.  However, it is not clear if
visualizing dissimilarities through the color channel is effective to
perform important tasks such as detection of clusters and outliers.


The background section describes the challenges of assessing
projections quality and explain how projection artifacts impact visual
analysis tasks based on projections.
We then introduce our method to visualize artifacts, and describe our
experiment design and its results before concluding.





%% file: tex/background.tex
\section{Background}

This section presents approaches that assess projections quality and tend to take projection artifacts into account.
To our knowledge, there is no controlled experiment to report on the effect
of these artifacts and their counter measures but we still report on
evaluations related to projections.

\subsection{Projections Quality}

Arguably, artifacts could be avoided or become negligible if the
projection quality were good enough.  Different approaches exist in
the literature to assess the quality of a dimensionality reduction
process~\cite{johansson2013quality}.  Because projection algorithms
are black-boxes, these approaches tend to help choosing and setting an
algorithm that will provide a projection that can be easily and
faithfully interpreted.  We can distinguish approaches that target an
automatic processing of dimensionality
reduction~\cite{bertini2011quality} and other interactive approaches
that provide measures and visualizations to help users explore
different projection settings, such as the DimStiller
system~\cite{ingram2010dimstiller}.  Both approaches have to define
first how to measure the quality of projections, and understanding the
measures is far from simple. 

The quality of projections can be defined through particular visual
criteria such as ``outlying'', ``clumpy'', ``skinny''~\cite{wilkinson2005graph};
it can also rely on task completion time benchmarks, such as visual
clustering~\cite{bertini2011quality}.  There are many ways to define
and evaluate visual clustering such as class consistency, cluster
separation~\cite{sips2009selecting} or class
density~\cite{tatu2010visual}.  However these automated measures still
do not compete with human judgment as it is hard to summarize visual
complexity and it is very dependent on data characteristics.
Moreover measures such as cluster separation were proved as not
reliable to judge the quality of a
projection~\cite{sedlmair2012taxonomy}.

A good projection mapping can be defined as one that minimizes the
\emph{projection stress}~\cite{torgerson1952multidimensional}, \ie the
least squares error between distances in data-space and
Euclidean distances in 2D space.  The closer the distances are, the
lower the stress is.  However, each projection algorithm, and
especially multidimensional scaling projections, gives its own stress
definition through the objective function it optimize.
Once a stress measure is defined, it can be visualized either
statically or interactively.

A common technique maps the stress measure optimized by the algorithm using a visual variable on each dots;
usually, each dot is colored depending on its stress value using a color scale from yellow (low stress) to red (high stress)~\cite{bentley1996animating}.
Other visual encoding techniques exist to display the stress measures such as \emph{jitter disc}~\cite{brodbeck1997domesticating}, or \emph{interpolated coloring}~\cite{schreck2010techniques}. 
However, visualizing the stress alone only warns users about potential
issues with the position of the points, but they do not provide any hint on
the nature of these issues. 

\subsection{Projection artifacts}

Even if projection algorithms tend to preserve ``most of'' the
important aspects of the underlying data structures existing in
data-space, all the 2D distances do not faithfully respect data-space
distances: some points are misplaced and we call these points
\emph{artifacts}.  We distinguish geometric
artifacts~\cite{aupetit2007visualizing} from topological
artifacts~\cite{lespinats2011checkviz}.  

Geometric artifacts are
caused by small distortions of distances such as compressions and
stretching.  The perceived 2D distances are inaccurate but
neighborhoods are preserved: these artifacts do not impact directly
interpretation trustworthiness as clusters are neither split nor merged. 
When distortions are ``important'', they can lead to topological artifacts:
the visible neighborhood is wrong, some items seem close when they are actually
far-away in data-space (false neighbors) while items that are
close in data-space become far-away in 2D-space (tears). 
These artifacts are more problematic for analysis tasks: they distort 2D
clustering and lead to misinterpretation.

\if\displayFigures0 \figureArtifacts \fi

Topological artifacts involve two important aspects: first, they are
\emph{relative} to a reference item or cluster.  For example, a point
that is a false neighbor for its 2D neighbors may also be a tear for
its data-space neighbors (\autoref{fig:distortions}).  Second,
topological artifacts have different levels of \emph{granularity}.
The definition of these artifacts relative to a reference item
transfers relative to a reference cluster.  Two distinct clusters in
data-space may be projected in the same area on the projection and
then become false neighbors.  In the same way, one cluster in
data-space can be split into disconnected components in 2D space, and
each component is a tear relative to the other.  Splitting clusters
``only'' impacts interpretation trustworthiness made from the
projection, but false neighbors also impact the visual quality of the
projection by deteriorating the visual separability of clusters.
These artifacts damage the interpretability of projections at
different levels depending on the considered visual analysis task.

Recently, Martins et al.~\cite{martins2014visual} proposed a system
with different views, each one indicating either the global stress, or
false neighbors or tears according to different scales (relative to an
item or a cluster).  Different measures are proposed for each type of
artifacts and are represented on each view using color interpolation
or bundled edges linking the original neighbors on the projection to
highlight tears. These edges are colored according to the error
intensity.  To control the clutter of the edges, user can select a
point or a set of points to focus the visualization around their
relative topological artifacts.  This system helps understand the
impact of the parameters and algorithm choices on the projection
quality, but it does not directly help to reveal actual clusters hidden by
projection artifacts.

\if\displayFigures0 \figureTasks \fi

Using a uniform color scale, the \emph{proximity-based visualization}~\cite{aupetit2007visualizing} corresponds to the interactive coloring of projections based on the original dissimilarities,
that we presented in \autoref{fig:tasks}.  This technique displays
interactively at each point its original dissimilarity in the data-space
relative to a reference data item selected by the user on the
projection. 
Distortions are indirectly visualized by contrast between 2D positions and colors
representing the original dissimilarities.  This technique may address
visual analysis tasks such as validating outliers or visual
clustering, but it displays local information relative to a reference
item and its visual encoding needs to be validated. 
In this paper we report on a controlled experiment with the aim of evaluating
the effectiveness of this technique compared to the bare projection with
regard to different visual analysis tasks. 

\subsection{Projections Evaluation}

Several kinds of studies exist to compare the performances of
projection algorithms ~\cite{biswas1981evaluation}.  These studies
consider a quality metric and a predefined setup of parameters to
compare the algorithms on benchmark data sets.  However, these studies
do not address the human ability to interpret the results of the
algorithms.

Few user studies have been done on how the visualized projection are
interpreted~\cite{icke2011automated}.  Recently, an evaluation on how
experts and novices judge the quality of projections has been
published~\cite{lewis2012behavioral}, showing that experts are
consistent in their analysis where novices are more random.  The
visual encoding of the projection as a 2D, 3D or SPLOM scatterplot has
also been studied~\cite{sedlmair2013empirical} and shows that the 2D
scatterplot used with SPLOM really helps user perform exploratory
analysis with projections.

%% file: tex/evaluation.tex
\section{Assessing the Impact of Projection Artifacts}

In this article, we want to assess how projection artifacts impact the
interpretation of multidimensional scaling techniques, and in
particular if we can overcome these artifacts using a ---preferably
simple--- technique.

The \emph{proximity-based visualization} \cite{aupetit2007visualizing}
is a simple technique;  unlike other techniques it does
not require any parameter, only the user selection of a reference
point.  The idea is to convey proximities in data-space using
color in addition to their spatial encoding.  Displayed dissimilarities
are relative to a reference item that implies exploring interactively
the projection using a pointer.  As this technique displays original
information in addition to the projection mapping, it can help dealing
with projection artifacts and help building a more reliable mental
model of the proximity relationships in data-space.  

However it is not clear if visualizing the dissimilarities using color is
effective to help performing visual analysis tasks on projections that
present artifacts.  Therefore, we designed a controlled experiment in
order to assess the potential of this technique.  With this
experiment, we address the following research questions:
\begin{enumerate}
\item Can we help non specialists deal with visual analysis tasks
  using projections, independently of the inherent quality of these
  projections?
\item How do artifacts impact the visual analysis of projections?
\item Is the interactivity required by proximity-based visualization
  worth its improvements in accuracy?
\end{enumerate}

In the following, we refer the bare projection simply as
\emph{Projection} and the \emph{proximity-based visualization}
\cite{aupetit2007visualizing} as \emph{ProxiViz}.  We first describe
the tasks, techniques and datasets we consider in the experiment before presenting
how we assess projection quality for each task. We then introduce the experiment procedure and
participants before presenting our hypotheses and the experiment results.

\if\displayFigures0 \figureDatasets \fi

\subsection{Tasks}

Recently, a characterization of task sequences for the visualization of dimensionality-reduced data has been established from interviews with data analysts \cite{brehmer2014visualizing}.
In this article, we consider multidimensional scaling techniques, therefore we focus only on cluster-oriented task sequences.
Among the three task sequences proposed, we consider only the sequences ``verify clusters'' and ``match clusters and classes'', as the sequence ``name clusters'' would require participants to have prior knowledge in data.
We define data outliers as particular cases of clusters: they are composed of only one data item; And class outliers are defined as mislabeled data items.
We considered the following tasks and questions:
\begin{description}[leftmargin=0px]
\item[Data outlier validation:] for a highlighted dot on the
  projection, participants have to choose a yes/no answer to the
  question ``Is this a data outlier?''.

\item[Clustering validation:] for two differently colored sets of
  highlighted dots on the projection, participants have to choose a
  yes/no answer to the question ``Do these two sets of points belong
  to the same cluster?''.

\item[Clusters enumeration:] participants have to answer the question
  ``How many clusters can you count?'' looking at the projection.
  They are allowed to answer 1 to 7.

\item[Class outliers validation:] for a highlighted dot on the class colored projection,
participants have to choose a yes/no answer to the question ``Is this a class outlier?''.
\end{description}

\subsection{Techniques}

We used the two following techniques:
\begin{description}[leftmargin=0px]
\item[\emph{Projection}] is static and displays a dot for each point
  with no color encoding.  This visualization corresponds to a spatial
  encoding of the dissimilarity matrix.  2D distances between points
  tend to respect the original dissimilarities between data items.

\item[\emph{ProxiViz}] is interactive and displays colors representing
  the original dissimilarities relative to a reference point.
  This interactive technique is a color encoding of only one row of the
  dissimilarity matrix.  When the mouse moves over the visualization
  area, the nearest point is interactively selected as
  \emph{reference}~\cite{aupetit2007visualizing}; 
  the color encoding of the visualization is immediately updated to display dissimilarities to the reference in the data-space.
  We use a uniform shade of blue colorscale that vary in intensity to color the proximities (\ie inverse \emph{dissimilarities}).  It starts with a white color indicating close-by items (\ie with a null dissimilarity),
  then continues with blue nuances, and ends with black, indicating
  the farest items with maximum dissimilarity to the reference.
\end{description}

\if\displayFigures0 \figureProxiViz \fi

The original \emph{proximity-based visualization} uses color to encode dissimilarities
and it applies color on the Voronoi cell of each point.  
However the size of the Voronoi cells depends on the density of points and may
vary arbitrary across the projection, which can introduce biases as this
size does not encode any information.

We performed a pilot experiment with both synthetic and real datasets
to choose a good visual encoding for ProxiViz. 
We compared applying coloring to points, to Voronoi cells and to background using color interpolation based on  the inverse-distance weighting.
This interpolation used with ProxiViz displays the global proximity trends in each area
of the projection while preserving the local proximity information with a shaded circle around each point (\autoref{fig:proxiviz}).

For a cluster enumeration task, the results of the pilot show that
ProxiViz with the color interpolation was significantly more accurate
that the other color encodings and overall preferred by the
participants for its precision and aesthetic.
We implemented the projection visualization in \verb|d3.js| \cite{bostock2011d3} and the color interpolation using WebGL Shaders.
This implementation scales up to one thousand items, so the interaction was fluid with our datasets.

\subsection{Datasets}

We used high dimensional datasets and made sure that the ground truth (corresponding to class
labels) was verified by the dissimilarity measure. 
More precisely, we verified that for the dissimilarities in the data-space, one
cluster exists for each class label and that this cluster does not
overlap with other clusters. We chose only datasets for which this criterion was verified.
We then projected datasets using two different projection algorithms:
classical MDS \cite{torgerson1952multidimensional}, 
erV \cite{venna2010information}, using different settings of the parameters.
We visually checked the clusters separability and selected the 
projections to balance questions difficulties.

Real datasets were mainly taken from \cite{UCI}: 
Pen-Based Recognition of Handwritten Digits (\emph{pen-digits}), 
Optical Recognition of Handwritten Digits (\emph{opt-digits}),
Isolated Letter Speech Recognition (\emph{isolet}),
Letter Recognition Dataset (\emph{letters}),
CMU Face Images Dataset (\emph{faces})
and Teapot Dataset (\emph{teapot}) from \cite{zhu2005harmonic}.
We also generated two synthetic datasets with clusters based on Gaussian distributions with random noise applied to some features of the second dataset (\emph{gaussians} and \emph{noisy gaussians}).

\if\displayFigures0 \figureDatasetsDetails \fi

In order to justify the use of multidimensional scaling instead of linear projections, we favored datasets from image and signal processing as their dimensions were not directly interpretable. 
The two synthetical datasets helped to balance questions difficulties.
More details are provided in \autoref{fig:dataset_summary}.

\subsection{Difficulty}

We visually explored each projection and detected local
artifacts. Then we set the questions and their answers manually for
each task and dataset in order to balance the questions in two
categories: easy and difficult.  Easy questions are questions that can
be answered with no additional information and difficult ones involve
topological artifacts or cluster separation problems.

We used visual exploration and automated algorithm to suggest clusters
and possible outliers using the scikit-learn library
\cite{pedregosa2011scikit}.  We then selected interesting cases
balancing the questions difficulties:
\begin{description}[leftmargin=0px]
\item[Data outlier validation:] we had to find data outliers for each
  dataset.  Using One Class SVM \cite{schlkopf1999support} and Local
  Outlier Factor \cite{breunig2000lof}, we selected two sets of data
  outliers.  We visually selected one outlier at the intersection of
  the two sets.  An easy question is when the outlier is well
  separated from the closest blob of points whereas a difficult one is
  when it is not.

\item[Clustering validation:] we had to select different clusters and
  cluster components for each dataset.  We visually selected two sets
  of points close in 2D space and we analyzed their connections using
  the k-nearest neighbors.  Difficult questions involve either tears
  or false neighborhoods.

\item[Clusters validation:] we had to analyze the quality of
  clustering of each dataset.  We verified the ground truth
  corresponding to labels by checking that each class had an
  underlying cluster and that clusters did not overlap too much each
  other.
  We then categorized the questions depending on how clusters were
  visually separated on the projection.

\item[Class outlier validation:] we had to extract class outliers for
  each dataset.  Using KNN Classifier \cite{cover2006nearest} and One
  Class SVM \cite{schlkopf1999support} on each class, we visually
  determined an interesting class outlier for each dataset.  Difficult
  questions involve either tears or false neighborhoods.
\end{description}

We tried to avoid questions that would be too easy or too hard to
answer either with Projection or ProxiViz.  We expected that ProxiViz
would be more accurate than Projection on difficult questions and that
the techniques would be equally accurate on easy questions.

\subsection{Experiment design}

We used a within-subjects design with no repeated measures and the
following factors: Technique and Difficulty.  The order of blocks and
trials were counterbalanced using a Latin Square.  We also changed the
orientation and flipped the display along one axis between each block
for each projection.  Each participant faced successively 2 blocks
(one for each technique) of 24 trials (3 tasks $\times$ 8 projections)
and then 2 blocks for the class outliers validation task with 8 trials
each.

We kept the blocks order with Projection first, because ProxiViz
displays more information and users would be biased when revisiting
the same projection using the bare condition. Indeed, during the pilot
session, we observed that participants recognized datasets when using
ProxiViz before Projection and not when using Projection first.
Introducing this order for the techniques may slightly bias our
results, but we believe that considering our high level tasks, it is
important to avoid the observed learning effect while keeping the
conditions equivalent by reusing the same projections.

In summary, the design included:
 \begin{center}
\begin{tabular}{l l l}
  4 & projections & $\times$ \\
  2 & difficulties & $\times$ \\
  4 & tasks & $\times$ \\
  2 & techniques (blocks) & = \\ 
  \hline 
  64 & trials per participants & $\times$ \\ 
  24 & participants & = \\ 
  \hline \textbf{1,536} & \textbf{trials in total} & \\
\end{tabular}
 \end{center}

\subsection{Participants}

The participants of our experiment are data analysts who aim at
understanding underlying data structures in high-dimensional data, \ie
outliers or clusters.  These data analysts can be domain experts,
statisticians, or data miners; they know how to interpret projections
resulting from dimensionality reduction and we made sure they were
aware of their possible artifacts.

24 participants (8 women and 16 men) from 23 to 37 years old (avg. 28)
from two laboratories specialized in data analysis completed the
evaluation.  They all had a computer science background but in
different domains: statistics, signal processing, image processing,
machine learning, data mining. Their application fields were also
varied 
in the same way that their level of expertise regarding data analysis. 
They rarely looked at projections 
but they had knowledge about projections techniques overall.
They were used to static visualizations such as those available in different statistical analysis software (matlab, R). 
They all had a normal vision with no color vision problems. 

We did not consider recruiting novice analysts as it would have
required training them to understand dimensionality reduction and
perform visual analysis of projections; still, our participants were not
experts in multidimensional scaling.

\subsection{Procedure}

We recorded the answer and response time of each participant for each
question.  Participants used only the mouse to interact with the system and answered questions by selecting a row in a multiple-choice form (yes/no or 1 to 7 clusters).
Timing was started once a
technique appeared on the screen and stopped when the participant
validated the answer.  After each validation, we also asked the
confidence level regarding the answer (five-points Likert scale).
Between each block, the participants were free to rest a
bit.  We imposed a duration of 30-60 seconds maximum for each trial,
for a total evaluation duration of around 30 minutes, preceded by 20
minutes of presentation of the evaluation and a training. The
evaluation system was run in a Chrome web-browser displayed in
full-screen resolution 1440 $\times$ 900 of a MacBook Pro
display. 

After a short introduction on artifacts and visual analysis, we
verified that the analysts understood the techniques and agreed with
our definitions of data outlier, cluster and class outlier.  
We trained subjects on a simple dataset for each technique and task, and we let them explain
their reasoning for each answer.  If necessary, we explained the
pitfalls of their reasoning and how they should understand the problem.

We used 6 subjects for a pilot to update the procedure of the
experiment and verify results consistency.  Before the study,
participants filled a questionnaire eliciting demographic information
and to check for possible color blindness.  After the study, participants
also filled a questionnaire of subjective preferences for the
different techniques to collect feedback on their user experience.

%% file: tex/results.tex
\subsection{Hypotheses}

Our hypotheses are that ProxiViz will be more accurate than Projection
overall, and even more so on difficult questions, both for local tasks
involving outliers or clusters, and visual clustering tasks implying
to explore the whole projection.  We also hypothesize that participants
will be significantly slower using the ProxiViz technique.  However we
do not expect significant differences on easy questions for each task
between ProxiViz and Projection.  

Based on our experience and our three research questions introduced
previously, our hypotheses were as follow:
\begin{description}
\item[H1] For each task, we expect that ProxiViz is more effective
  than Projection in terms of accuracy.

\item[H2] For each task, we expect ProxiViz to be more effective than
  Projection for difficult questions and as effective as Projection
  for easy questions.

\item[H3] For each task, we expect that participants will be
  significantly faster using Projection.
\end{description}

\section{Results}

\if\displayFigures0 \figureResultsCorrectness \fi
\if\displayFigures0 \figureResultsTime\fi

We transformed the error into a percentage of correctness: for tasks
that expect a yes/no response, the correctness is either 100\% or 0\%
if the answer is right or wrong.  For clusters enumeration, the error
is calculated as the difference between the number of clusters
found $c$ and the number of clusters really present in the data $c^*$.
We use in the sequel a percentage of correctness $v$ based on this formula:

\begin{equation}
  \label{correctness}
  v = 
 \left\{
 \begin{array}{l}
   |\frac{c - c^*}{8 - c^*}| \times 100 \mbox{, if } c > c^* \\
   |\frac{c - c^*}{1 - c^*}| \times 100 \mbox{, if } c \leq c^*
 \end{array}
 \right.
\end{equation}

Since our design is within-subject with repeated measures and that
correctness do not follow a normal distribution for all tasks, we used
Friedman's test for one way analysis of variance to analyze more than
two non-parametric distributions, and pairwise Wilcoxon signed rank
test with Bonferroni correction to analyze two independent
non-parametric distributions.  We log-transformed the response time to
better fit a normal distribution and verified this assumption using a
Shapiro-Wilk test ($p > .05$).  We then used the one-way repeated
measure ANOVA and paired t-tests to analyze response time.

In terms of correctness, Friedman's test reveals a significant effect
of the factor Technique for each task, except for clusters
enumeration. A significant effect of both Technique and Difficulty was
found for each task.  Overall participants obtained more good answers
using ProxiViz than Projection for each task except for clusters
enumeration.  The details are shown in \autoref{fig:results1}.

In terms of response time, ANOVA reveals a significant effect of the
factor Technique and no significant interaction between Technique and
Difficulty.  No significant differences were found in response time
between easy and difficult questions for each pair of technique and
task.  Pairwise comparison of the two techniques revealed that
ProxiViz is significantly around 2 times slower ($p < .0001$) than
Projection for each task, except clusters enumeration for which
ProxiViz is significantly 3 times slower ($p < .0001$) and both on
easy and difficult questions.  More details are shown in
\autoref{fig:results2}.

\subsection{Confidence and User Preferences}

No significant differences were found for the confidence answers
between the two techniques.  For all tasks, users felt confident
except for clusters enumeration where they were neither confident nor
skeptical for both technique.  Low correlations were found between
confidence and correctness, and between confidence and time.

We used these confidence results to check that difficulty levels were
well calibrated and that none of the questions were too difficult or
too easy to answer from the participants' point of view.  We would have
expected that participants felt more confident using ProxiViz but it
was not the case as they were not always able to interpret easily the
coloring or to reconstruct the clustering using relative proximities gathered
during exploration.

Overall ProxiViz was preferred to Projection (20 participants).  18
participants felt more confident and preferred ProxiViz for its
clarity.  18 participants felt faster using Projection.  21 felt that they
were making correct choices using ProxiViz and 16 were satisfied by
their choices.  Some participants said they felt responsible for the
errors because they were not able to accurately decode the additional
information provided by ProxiViz.

%% file: tex/discussion.tex
\section{Discussion}

In this section, we discuss the results regarding 
our hypotheses which are overall supported.

\medskip

\noindent\textbf{H1: For each task, we expect that ProxiViz is more effective
  than Projection in terms of accuracy.}

For data and class outliers validation, \textbf{H1} is supported as
overall, ProxiViz was significantly (16\%) more accurate than
Projection.  For clustering validation also, ProxiViz is significantly
(35\%) more accurate than Projection.  However, \textbf{H1} is
partially supported because for clusters enumeration ProxiViz is not
significantly more accurate than Projection.

This last result is not surprising as ProxiViz gives better insights
on local information relative to one reference item.  Generalizing
each relative coloring to perform visual clustering requires
to explore the whole projection with the pointer, relying on the human
memory to reconstruct the clustering, which implies a high cognitive
load. 
In particular, we noticed during the evaluation that participants felt
lost on this task using ProxiViz as the dissimilarity distributions were
really different from one dataset to the other.  Some participants felt
also lost during their exploration as colors changed dramatically when
moving from one reference to another.

Moreover the difference of color intensities between clusters was
sometimes difficult to assess to delimit clusters.  In particular, we
noticed that some participants used the 2D positions instead of the
coloring when they considered that color difference was too difficult
to interpret and the 2D distances good enough.  This suggests that the
bare projection gives a suitable overview of the data clustering.

Overall results show that participants were on average effective at
analyzing the colors displayed by ProxiViz (60\% to 80\% accurate).
ProxiViz is also significantly more reliable than Projection for
outliers and clusters validation.  So ProxiViz can help
non-specialists deal with label matching tasks and tasks related to
class structure in a confirmatory context, irrespective of the
inherent quality of the projection considered.  Note that we
considered a dissimilarity matrix that reflected the class model, and
we did not took into account the inherent problem of finding and
validating such dissimilarity measure.

\medskip

\noindent\textbf{H2: For each task, we expect ProxiViz to be more effective than
  Projection for difficult questions and as effective as Projection
  for easy questions.}

For data and class outliers validation, \textbf{H2} was supported as
ProxiViz was 30-38\% more accurate than Projection on difficult
questions.  For clustering validation also as ProxiViz was 74\% more
accurate than Projection on difficult questions.  Furthermore, for
class outlier validation, we observed that the 6\% difference of
correctness on easy questions between ProxiViz and Projection was
significant.

For clustering validation, correctness results are very low for hard
questions using Projection.  For this task, artifacts involved in hard
questions were almost impossible to detect using Projection.  However
using ProxiViz, they were revealed and participants were able to use
their finding to answer questions without having to explore, which
explains that correctess results were better for hard questions than
for easy questions using ProxiViz.

We noticed during the evaluation that participants had sometimes
problems to recall the definition of a class outlier and the question
may have not been well expressed.  For this task, participants had to
verify that the highlighted point was not in the neighborhood of the
other points of its class.  So they had to verify that: ``no, the
point is not in the neighborhood of the class'' to answer ``yes, this
is a class outlier''.  Using Projection, verifying the neighborhood is
quickly done and some participants may have made mistakes because they
answered too quickly, not following the right reasoning, which was
clearly uneasy.

For clusters enumeration, \textbf{H2} was partially supported, as
ProxiViz was not less accurate than Projection on difficult questions.
Nevertheless for clusters enumeration, we observed that the 6\%
difference of correctness on easy questions between Projection and
ProxiViz was significant.  This suggest that when clusters are well
separated, Projection gives a better overview of the data
clustering than ProxiViz.

Considering all tasks and techniques, we can notice that the highest
correctness is close to 90\% and the lowest correctness close to 5\%.
This suggests that questions were neither too easy nor too hard, and
especially for local tasks that required a yes/no answer.  These
results might be improved using a longer training of participants
to make them share the same visual criteria defining outliers and
clusters.

Overall, our results show that ProxiViz is more robust to artifacts
than Projection.  For local tasks, analysts must clearly be
explained the dangerous impact of artifacts.  Conversely for more
global tasks involving visual clustering, it seems that the impact of
artifacts is less important.  The static projection is a good tradeoff
to obtain an overview of the data clustering.  However we still need
to evaluate if other techniques that enhance the projection may
improve the accuracy of visual clustering.

\medskip

\noindent\textbf{H3: For each task, we expect that participants will be
  significantly faster using Projection}

For data outliers, class outliers and clusters validation, \textbf{H3}
was supported, as overall ProxiViz was significantly slower (around 2
times) than Projection on both easy and difficult questions.  For
clusters enumeration, \textbf{H3} was also supported, as overall,
ProxiViz was significantly slower (around 3 times) than Projection on both
easy and difficult questions.  Overall no significant differences were
found between easy and difficult questions for each technique and
task.

Exploring projections for local tasks with ProxiViz does not
require a large difference of time compared with Projection.
Moreover this time is independent of artifacts. 
ProxiViz is then worthwhile to dig into the details, as
users will feel more responsible for their choices because these choices
are resulting from a methodical exploration instead of an overview for
which we know artifacts happen.

%% file: tex/conclusion.tex
\section{Conclusion and Future Work}

In this article, we have investigated how artifacts inherent to
multidimensional projections interfere with the interpretation of
these visualizations, and the effectiveness of the ProxiViz technique
to facilitate the detection and interpretation of these artifacts.
The ProxiViz technique interactively colors the whole visualization
when the user moves the pointer to a particular point: it shows a
color-map where the intensity decreases with the dissimilarity in
data-space from the focus point, conveying dissimilarity information using
the color channel.  We conducted a controlled experiment to evaluate
the effectiveness of ProxiViz compared to the bare projection with
regard to different visual analysis tasks. These tasks require both
local and global exploration, and involve both outliers and clusters
detection. 

Results show that ProxiViz is effective to perform local tasks such as
label matching of outliers and clusters.  Moreover this interactive
technique is robust to the artifacts whereas the bare projection is
not for the considered tasks.  Conversely, for visual clustering
tasks, results show that the impact of artifacts is less important and
that projections give a suitable overview of the data clustering.  The
ProxiViz technique is therefore a valuable interactive enhancement for
projection-based systems.

Projection-based visualizations have applications in domains where users
are not analysts but need to make important decisions about
high-dimensional data.  
One example is cargo scanning, used by customs
to check the contents of freight shipping containers as quickly as
possible.  The scan of a container returns multidimensional
information that needs to be checked against the list of goods
declared to be in the container.  This list of goods (\eg toys,
furniture) is translated into multidimensional information and
projected over the scanned information.  Some decision needs to be
done about the container: let it go, fast check, or in-depth check.
This decision can be critical and relies on human judgment to
interpret the possible issues.  Our results indicate that ProxiViz is
suitable for this use case as is robust to projection
artifacts for outlier detection.

As projection artifacts have still a significant effect on local
analysis, measures and approaches that help assessing the quality of
projection must be generalized in existing softwares that visualize
high-dimensional data with projections.